\documentclass[aps,prc,twocolumn,superscriptaddress,showpacs,showkeys]{revtex4}
\usepackage{amsmath,amssymb}
\usepackage{bm}
\usepackage[dvips]{graphicx}
\newcommand{\re}{\text{Re}}

\begin{document}

\title{Study of $\Lambda(1405)$ in photoproduction of $K^*$}

\author{T.~Hyodo}%
\affiliation{Research Center for Nuclear Physics (RCNP), 
Ibaraki, Osaka 567-0047, Japan}%

\author{A.~Hosaka}%
\affiliation{Research Center for Nuclear Physics (RCNP), 
Ibaraki, Osaka 567-0047, Japan}%

\author{M.~J.~Vicente~Vacas}%
\affiliation{Departmento de F\'isica Te\'orica and IFIC,
Centro Mixto Universidad de Valencia-CSIC,
Institutos de Investigaci\'on de Paterna, Aptd. 22085, 46071
Valencia, Spain}%

\author{E.~Oset}%
\affiliation{Departmento de F\'isica Te\'orica and IFIC,
Centro Mixto Universidad de Valencia-CSIC,
Institutos de Investigaci\'on de Paterna, Aptd. 22085, 46071
Valencia, Spain}%

\date{\today}

\begin{abstract}
    We investigate the photoproduction of $K^*$ vector meson 
    for the study of the $\Lambda(1405)$ resonance.   
    The invariant mass distribution of $\pi\Sigma$ shows a
    different shape from the nominal one, peaking at 1420 MeV.
    This is considered as a consequence of the double pole
    structure of $\Lambda(1405)$, predicted in the chiral
    unitary model.
    Combined with other reactions, such as $\pi^- p \to K^0 \pi
    \Sigma$,
    experimental confirmation of this fact will reveal a novel
    structure of the $\Lambda(1405)$ state.
\end{abstract}


\maketitle

The structure of the $\Lambda(1405)$ resonance has 
been a long-standing problem in
hadron physics.
As is well known, 
it is not easy to derive a proper mass
for $\Lambda(1405)$ in a naive quark model,
while unitary coupled channel
approaches~\cite{annphys10.307,PR153.1617,Jones:1977yk},
in which the resonance is described as a quasi-bound state
of a meson and a baryon,
have been successful.
In recent years, 
these issues are reconsidered on the basis of chiral
effective Lagrangians and fundamental QCD.
The chiral unitary model~\cite{Kaiser:1995eg,Kaiser:1997js,
Oset:1998it,Lutz:2001yb}, which implements the interaction
derived from chiral perturbation theory and coupled channel
unitarity, has been well reproducing the $S=-1$
meson-baryon scattering amplitude, generating
$\Lambda(1405)$ dynamically.
One of the lattice QCD calculations has reported 
the difficulty in describing $\Lambda(1405)$ by three-quark interpolating 
operator~\cite{Nemoto:2003ft}.
These facts seem to favor the meson-baryon picture of the $\Lambda(1405)$
state, rather than a simple 3-quark state.

In this study, we focus on the pole structure of the 
$\Lambda(1405)$ resonance; 
it has been found that there are two poles 
in the region of $\Lambda(1405)$ through analyses based on 
the chiral unitary models~\cite{Oller:2000fj,Oset:2001cn,Jido:2002yz,
Garcia-Recio:2002td,
Hyodo:2002pk,Hyodo:2003qa,Garcia-Recio:2003ks,Nam:2003ch}.  
The conclusion does not depend on the details of the models,
qualitatively.
The existence of two poles was first found in the 
cloudy bag model~\cite{Fink:1990uk}.
Very recently, a study of $1/2^-$ pentaquark
states with correlated diquark picture~\cite{Zhang:2004xt} 
also reported two $\Lambda$ states at relatively low energy region.

The detailed structure of these poles have been studied
in the chiral unitary model, which we use in this study.
For instance, in Ref.~\cite{Jido:2003cb}, the positions of the poles 
and its coupling strengths to meson-baryon channels
are calculated as in Table.~\ref{tbl:poles}.
We see that the pole $z_1$
couples dominantly to $\pi \Sigma$ channels, 
while $z_2$
couples dominantly to $\bar K N$ channels.  
Under this situation, the shape of $\Lambda(1405)$
seen in invariant mass distribution of $\pi\Sigma$
should depend on the reaction to generate the resonance.
In fact, such differences were seen in previous theoretical 
studies~\cite{Nacher:1998mi,Nacher:1999ni,Hyodo:2003jw}.
It was found in Ref.~\cite{Hyodo:2003jw} that
the $z_1$ pole was favored in the $\pi^- p \to K^0 \pi \Sigma$ reaction.

\begin{table}[tbp]
    \centering
    \caption{Pole positions and coupling strengths $g_i$ for several 
    channels in the chiral
    unitary model~\cite{Jido:2003cb}.}
    \begin{ruledtabular}
    \begin{tabular}{ccccc}
	pole &  $\pi\Sigma$ & $\bar{K}N$ & $\eta \Lambda$ & $K\Xi$ \\
	\hline
	$z_1 = 1390 - 66i$ & 2.9 & 2.1
	& 0.77 & 0.61   \\
	$z_2 = 1426 - 16i$ & 1.5 & 2.7
	& 1.4 & 0.35   \\
    \end{tabular}
    \end{ruledtabular}
    \label{tbl:poles}
\end{table}

\begin{figure}[tbp]
    \centering
    \includegraphics[width=8cm,clip]{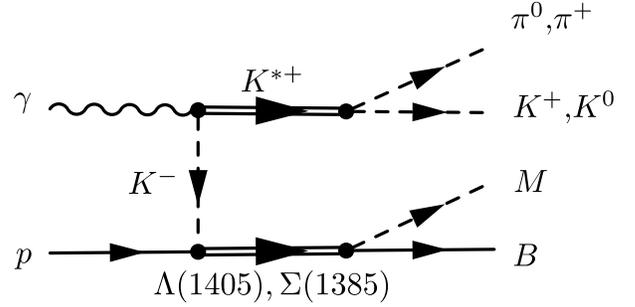}
    \caption{\label{Lambdareaction}
    Feynman diagram for the reaction.  
    $M$ and $B$ denote
    the meson and baryon of
    ten coupled channels of $S=-1$ meson-baryon scattering.
    In this paper we only take $\pi\Sigma$ and $\pi\Lambda$ channels
    into account.}
\end{figure}

In this paper, we propose
$\gamma p \to K^* \Lambda(1405)\to \pi^+ K^0 MB$ reaction
in order to isolate the $z_2$ pole.  
The advantages of this reaction are
\begin{itemize}
    \item  In $K$ exchange diagram as shown in
    Fig.~\ref{Lambdareaction},
    the pole $z_2$ will be selected, because
    $\bar{K}N$ channel couples to $\Lambda(1405)$
    at initial stage.

    \item  We can select the above events, using
    the correlation between polarization of photon beam
    and angular distribution of final $\pi^+ K^0$~\cite{Hyodo:2004vt}.
\end{itemize}
In the present calculation, we consider the threshold production
of $K^*$ and $\Lambda(1405)$, and utilize the $s$-wave
meson-baryon scattering amplitude calculated by the chiral unitary
model~\cite{Oset:2001cn,Hyodo:2002pk}, as the final state interaction
of $\bar{K}N$. In order to perform a realistic calculation,
we introduce the $p$-wave $\Sigma(1385)$ field explicitly.

The scattering amplitude as described by the 
diagram of Fig.~\ref{Lambdareaction} can be divided into two
parts
\begin{equation}
    -it=(-it_{\gamma\to K^-K\pi})\frac{i}{p_{K^-}^2-m_{K^-}^2}
    (-it_{K^-p\to MB}) \ .
    \label{eq:totamp}
\end{equation}
The former part $(-it_{\gamma\to K^-K\pi})$,
is derived from the 
effective Lagrangians~\cite{Hyodo:2004vt},
and given as
\begin{equation}
    -it_{\gamma\to K^-K^0\pi^+}
    = \frac{i
    \sqrt{2}g_{VPP}
    \epsilon^{\mu\nu\alpha\beta}p_{\mu}(K^0)p_{\nu}(\pi^+)
    k_{\alpha}(\gamma)\epsilon_{\beta}}
    {P_{K^*}^2-M_{K^*}^2+iM_{K^*}\Gamma_{K^*}}
    \ , 
    \label{eq:vecamp}
\end{equation}
where $p$ and $k$ are the momenta of the particle in parentheses,
$\epsilon_{\mu}$ the polarization vector of photon,
$g_{VPP}=-6.05$ the universal vector meson coupling, and
$\Gamma_{K^*}$ the total decay width of $K^*$, for which 
we take the energy dependence into account.
It is easy to see the $\pi^+ K^0$ distribution is correlated with
the polarization of initial photon.

\begin{figure}[tbp]
    \centering
    \includegraphics[width=8cm,clip]{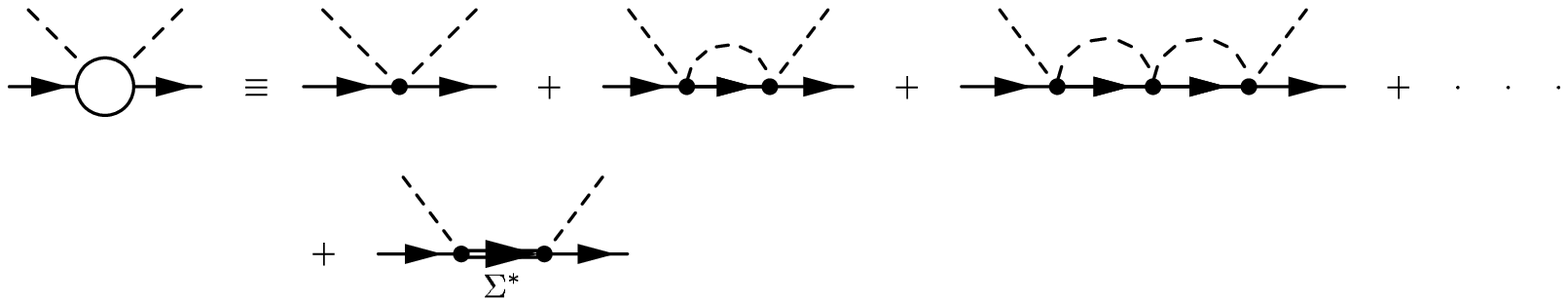}
    \caption{\label{fig:Baryon}
    Feynman diagram for $ K^- p \to MB$.}
\end{figure}

The amplitude $(-it_{K^-p\to MB})$,
shown in Fig.~\ref{fig:Baryon},
consists of two parts
\begin{equation}
    -it_{K^-p\to MB}(M_I)
    =-it_{ChU}(M_I)-it_{\Sigma^*}(M_I) \ ,
    \label{eq:baryonamp}
\end{equation}
where $-it_{ChU}$ is the meson-baryon scattering amplitude
derived from the chiral unitary model, and $-it_{\Sigma^*}$
is the $\Sigma(1385)$ pole term.
$M_I$ is the invariant mass for 
$K^-p$ system, which is determined by 
$M_I^2=(p_\gamma+p_N-p_{K^*})^2 $.
In the chiral unitary model~\cite{Oset:2001cn,Hyodo:2002pk}, 
the coupled channel amplitudes are obtained by
\begin{equation}
    t_{ChU}(M_I)=[1-VG]^{-1}V\, , 
    \label{eq:ChUamp}
\end{equation}
where $G$ is the meson-baryon loop function and $V$ is the
kernel interaction derived from the Weinberg-Tomozawa term
of the chiral Lagrangian. 
The $\Sigma(1385)$ term is introduced with
couplings $c_i$ to channel $i$ ($\Sigma(1385) \to MB$) 
which are deduced from the $\pi N \Delta$ using $SU(6)$ symmetry,
and explicit values are shown in Ref~\cite{Hyodo:2004vt}.
Then we have the amplitude
\begin{equation}
\begin{split}
    -it_{\Sigma^*}(M_I)
	=&- c_1c_i
	\left(\frac{12}{5}\frac{g_A}{2f}\right)^2
	\bm{S}\cdot \bm{k}_1
	\bm{S}^{\dag}\cdot \bm{k}_i \\
	&\times\frac{i}{M_I-M_{\Sigma^*}+i\Gamma_{\Sigma^*}/2}
	F_f(k_1) \ ,
\end{split}
    \label{eq:Sigmaamp}
\end{equation}
where $g_A=1.26$, $f=93\times 1.123$ MeV~\cite{Oset:2001cn}
and $\bm{S}$ is a spin transition operator.
We have introduced a strong form factor 
$F_f(k_1)= (\Lambda^2 - m_K^2)/(\Lambda^2 - k_1^2)$  with 
$\Lambda=1$ GeV,
for the vertex $K^-p\Sigma^*$
in order to account for the finite size structure of the baryons.

The cross section is given as a function of the incident energy $\sqrt{s}$:
\begin{equation}
    \begin{split}
	\sigma(\sqrt{s})
	    =&\frac{2MM_{\Sigma}}{s-M^2}
	    \int \frac{d^3p_1}{(2\pi)^3}\frac{1}{2\omega_1}
	    \int \frac{d^3p_2}{(2\pi)^3}\frac{1}{2\omega_2} \\
	    &\times\frac{1}{2}\int_{-1}^{1}d\cos\theta
	    \frac{1}{4\pi}\frac{\tilde{P}_3}{M_I} |t(\cos\theta)|^2
	    \ ,
    \end{split}
    \label{eq:cross}
\end{equation}
where $p_{1(2)}$ and $\omega_{1(2)}$ are the momenta and energy
of the final $K(\pi)$ from $K^*$, and $\tilde{P}_3$ is the relative 
three momentum of $MB$ ($\sim \pi \Sigma$ or $\pi \Lambda$) 
in their center of mass frame.  
The angle $\theta$ denotes the relative angle of $MB$ in the CM frame 
of the total system.   
The calculation is performed by the Monte-Carlo method.

Before going to the numerical results,
here we mention the $MB$ channels decaying from 
the intermediate baryonic states ($B^* \sim \Lambda(1405),
\Sigma(1385)$), below the threshold of the $\bar{K}N$ channel.
In the present case, since we have the $K^-p$ channel initially,
the $I=2$ component of $\pi\Sigma$ channel is not allowed,
and hence, there are two charged and two neutral channels.  
Considering the Clebsh-Gordan coefficients~\cite{Nacher:1998mi},
the charged channels ($\pi^{\pm}\Sigma^{\mp}$) are from the decay 
of both $\Lambda(1405) (I = 0)$
and $\Sigma(1385) (I = 1)$, while the neutral channels are 
from either one of the two;  
$\pi^0 \Sigma^0$ is from $\Lambda(1405)$ and 
$\pi^0 \Lambda$ is from $\Sigma(1385)$.

\begin{figure}[tbp]
    \centering
    \includegraphics[width=8cm,clip]{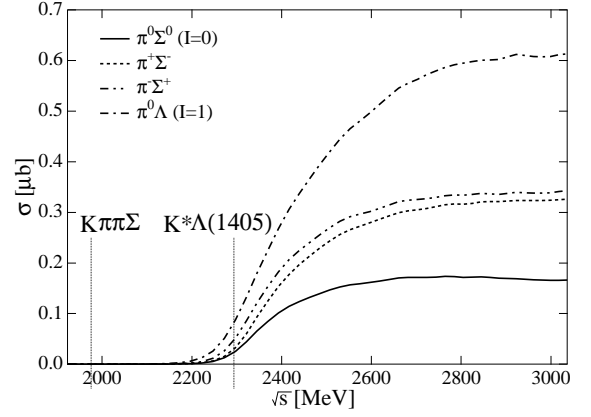}
    \caption{Total cross sections of the process
    with the final states $\pi^0\Sigma^0$ (Solid),
    $\pi^{+}\Sigma^{-}$ (Dashed),
    $\pi^{-}\Sigma^{+}$ (Dash-dot-dotted)
    and $\pi^0\Lambda$ (Dash-dotted)
    in units of [$\mu$b].
    Solid bars indicate the threshold energy of channels.}
    \label{fig:cross}
\end{figure}

In Fig.~\ref{fig:cross}, we show the total cross sections
$\sigma(\gamma p \to K^* B^* \to \pi^+K^0 MB)$ as 
functions of $\sqrt{s}$ for different $MB$ channels.  
The use of polarized beam enables us to reduce 
possible backgrounds,
but there is not distinction between cross sections of 
polarized and unpolarized processes,
unless we observe angular distributions.
In the figure, $\pi\Lambda(I=1)$ channel gives the largest magnitude,
which might disturb the $I=0$ amplitude, that we are interested in.
However, as we see below,
it is possible to isolate $\Lambda(1405)$ from $\Sigma(1385)$
contribution, with proper combination of final states.

\begin{figure}[tbp]
    \centering
    \includegraphics[width=8cm,clip]{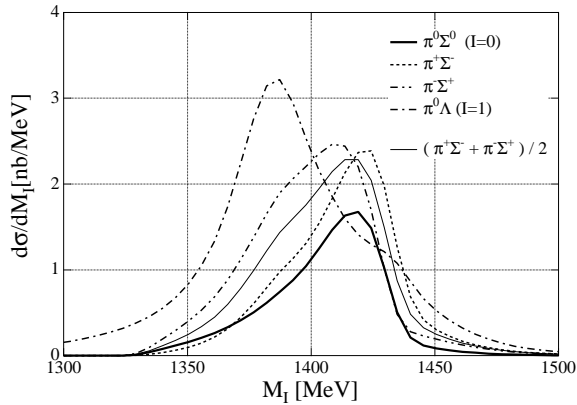}
    \caption{Invariant mass distributions
    of $\pi^0\Sigma^0$ (Thick solid),
    $\pi^{+}\Sigma^{-}$ (Dashed),
    $\pi^{-}\Sigma^{+}$ (Dash-dot-dotted),
    $\pi^0\Lambda$ (Dash-dotted)
    and $(\pi^{+}\Sigma^{-}+\pi^{+}\Sigma^{-})/2$ 
    (Thin solid)
    in units of [nb/MeV].
    Initial photon energy in Lab. frame is 2500 MeV 
    ($\sqrt{s}\sim
    2350$ MeV, 
    threshold of $K^*\Lambda(1405)$).}
    \label{fig:Mdist}
\end{figure}

In Fig.~\ref{fig:Mdist},
we show the invariant mass distributions for different decay channels
with  photon energy 
$E_\gamma = 2500$ MeV (the threshold for $K^* \Lambda(1405)$ 
production is 2350 MeV).  
As expected, the $\pi^0 \Sigma^0$ distribution decaying 
from $\Lambda(1405)$ (Thick solid line) has a peak 
around 1420 MeV which is the position of the $z_2$ pole.  
The $\pi^0 \Lambda$ distribution (dot-dashed line) 
has clearly 
a peak around 1385 MeV.
Forgetting about the experimental feasibility, these neutral channels 
are most helpful in order to distinguish the 
contributions from $\Lambda(1405)$ and $\Sigma(1385)$,
since they are pure $I=0$ or $1$.
In experiments, the charged states
(dashed and dash-dot-dotted lines)
may be observed, which 
contain both $\Lambda(1405)$ and $\Sigma(1385)$ contributions.  
The shapes of the three $\pi\Sigma$ distributions have a similar
tendency as the Kaon photoproduction process~\cite{Nacher:1998mi},
which has been confirmed in experiments~\cite{Ahn:2003mv}.
Note also that the contributions from $\Sigma(1385)$ seem to be small
for these channels.

It is worth showing the isospin
decomposition of the distributions of charged states~\cite{Nacher:1998mi}
\begin{equation}
    \begin{split}
	\frac{d\sigma(\pi^{\pm}\Sigma^{\mp})}{dM_I}
	&\propto \frac{1}{3}|T^{(0)}|^2+\frac{1}{2}|T^{(1)}|^2
	\pm\frac{2}{\sqrt{6}}\re (T^{(0)}T^{(1)*}) \ ; \\
	\frac{d\sigma(\pi^{0}\Sigma^{0})}{dM_I}
	&\propto \frac{1}{3}|T^{(0)}|^2 \ ,
    \end{split}
    \label{eq:isodecomp}
\end{equation}
where $T^{(I)}$ is the amplitude with isospin $I$.
This equation tells us that the difference between 
$\pi^+\Sigma^-$ and $\pi^-\Sigma^+$ 
comes from the crossed term $\re (T^{(0)}T^{(1)*})$,
and when we sum up the two distributions, this term vanishes.
In Fig,~\ref{fig:Mdist}, we show the result of average of 
the charged $\pi\Sigma$ channels (thin solid line), in order to remove
the contribution from the crossed term.
The peak of the distribution is still at around 1420 MeV,
because
the initial $K^- p$ couples dominantly to the second 
pole of the $\Lambda(1405)$, although the width of this distribution is slightly larger 
than that of the $I=0$ resonance due to a finite contribution 
from the $\Sigma(1385)$.

\begin{figure}[tbp]
    \centering
    \includegraphics[width=8cm,clip]{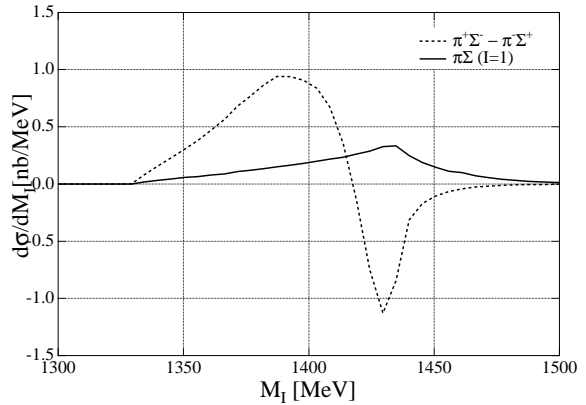}
    \caption{
    Invariant mass distributions
    of $\pi^{+}\Sigma^{-}-\pi^{-}\Sigma^{+}$ (Dashed),
    and $s$-wave, $\pi\Sigma(I=1)$ (Solid),
    in units of [nb/MeV].}
    \label{fig:I0amp}
\end{figure}

In the chiral unitary model,
$I=1$ $s$-wave amplitude has an interesting feature:
another pole is found at $1410 - 40i$~\cite{Oller:2000fj,Jido:2003cb}.
However, the existence of this pole is sensitive to the details of
the model, and in some cases, it appears in unphysical Riemann sheet.
But in all cases, the reflection of the pole could be seen on
the scattering line.
Usually, the $I=1$ amplitude can be extracted
by combining the three $\pi\Sigma$
channels:
\begin{equation}
    \frac{d\sigma(\pi^+\Sigma^-)}{dM_I}
    +\frac{d\sigma(\pi^-\Sigma^+)}{dM_I}    
    -2\frac{d\sigma(\pi^0\Sigma^0)}{dM_I}
    \propto|T^{(1)}|^2 \ .
     \label{eq:diffmanolo}
\end{equation}
However, the $|T^{(1)}|^2$ term 
would contain contributions both from $s$ and $p$-wave, although the
contribution of the $p$-wave to the $\pi\Sigma$ channels is small.
In order to extract the  $I=1$ $s$-wave amplitude,
we can utilize the crossed term in Eq.~\eqref{eq:isodecomp}.
Since we are looking at the cross sections where the angle
variable among $MB$ is integrated, the product of
$s$- and $p$-wave
amplitude vanishes.
Then, the difference of the distributions for the  two charged states 
contains only the $s$-wave $T^{(1)}$ amplitude
\begin{equation}
    \frac{d\sigma(\pi^+\Sigma^-)}{dM_I}
    -\frac{d\sigma(\pi^-\Sigma^+)}{dM_I}
    =\frac{4}{\sqrt{6}}\re (T^{(0)}_s
    (T^{(1)}_s)^*) \ .
    \label{eq:diff}
\end{equation}
We plot this magnitude in Fig.~\ref{fig:I0amp} with a dashed line.
In principle, it is possible to extract $T^{(1)}_s$ from this quantity
combining the distribution of $s$-wave $I=0$ (for instance, from the $\pi^0
\Sigma^0$).
Theoretically, in the present framework, we can calculate the pure
$s$-wave $I=1$ component
by switching off the $\Sigma(1385)$ and making the
combination of $\pi \Sigma$ amplitudes
as in Eq.~\eqref{eq:diffmanolo}.
The results
are shown in Fig.~\ref{fig:I0amp} (Solid line) and a small peak is seen as
a reflection of the approximate resonant structure  predicted in 
Refs.~\cite{Oller:2000fj,Jido:2003cb}.

In summary,
we have proposed a reaction 
$\gamma p \to \pi^+ K^0 MB$ for the study of 
the second pole possibly existing in the $\Lambda (1405)$ 
region.  
As expected, the mass distribution peaks at around 1420
MeV with a relatively narrow width,
and the present reaction is suitable for the isolation
of the second pole.
We have also shown that the effect of $\Sigma(1385)$
resonance for the $\pi\Sigma$ states is small enough
to see the structure of second pole of $\Lambda(1405)$.
The observation of different shapes of mass distribution
from the nominal one, such as in
$\pi^- p \to K^0 \pi\Sigma$~\cite{Thomas:1973uh},
can be the proof of the reflection of double pole structure.
Experimental evidence on the existence of such two
$\Lambda^*$ states would provide more information
on the nature of the current $\Lambda(1405)$ and thus new clues to 
understand non-perturbative dynamics of QCD.  
For more details, see Ref.~\cite{Hyodo:2004vt}.

\section*{Acknowledgments}
We would like to thank Prof. T.~Nakano for suggesting this work to us.
We also thank  Dr. D. Jido for useful comments and discussions.
This work is supported by the Japan-Europe (Spain) Research
Cooperation Program of Japan Society for the Promotion of Science
(JSPS) and Spanish Council for Scientific Research (CSIC), which
enabled T. H. and A. H. to visit
IFIC, Valencia and M.J.V. V. visit RCNP, Osaka.
This work is also supported in part  by DGICYT
projects BFM2000-1326,
and the EU network EURIDICE contract
HPRN-CT-2002-00311.

\end{document}